\documentclass[%
  english,
  biblatex,
  final,
]{lni}

\usepackage{bchart}
\usepackage{tabularx}
\usepackage{csquotes}
\usepackage{siunitx}
\usepackage{booktabs} 
\usepackage{fontawesome}
\usepackage{xcolor,colortbl}
\usepackage[inline]{enumitem}
\usepackage[
  colorinlistoftodos,
  textsize=tiny,
  obeyFinal,
]{todonotes}

\usepackage{xparse}
\NewDocumentCommand{\rot}{O{45} O{1em} m}{\makebox[#2][l]{\rotatebox{#1}{#3}}}%

\definecolor{Lightgray}{gray}{0.95}

\newcommand\rd{rough date}
\newcommand\od{ordering date}
\newcommand\vd{vanishing date}
\newcommand\rdf{\lstinline|RoughDateField|}
\newcommand\odf{\lstinline|OrderingDateField|}
\newcommand\vdf{\lstinline|VanishingDateField|}

\newcommand\vdt{\lstinline|VanishingDateTime|}
\newcommand\vp{\lstinline|ReductionPolicy|}
\newcommand\ve{\lstinline|ReductionEvent|}
\newcommand\mix{\lstinline|VanishingDateMixin|}

\newcommand\con{\lstinline|OrderingContext|}
\newcommand\dtf{\lstinline|DateTimeField|}

\begin{document}
\title{PrivacyDates: A Framework for More Privacy-Preserving Timestamp Data Types}

\author[%
  Christian Burkert
  \and Jonathan Balack
  \and Hannes Federrath
]{%
  Christian Burkert\footnote{%
    Universität Hamburg, \email{christian.burkert@uni-hamburg.de}
  }
  \and Jonathan Balack\footnote{%
    Universität Hamburg, \email{jonathan.balack@informatik.uni-hamburg.de}
  }
  \and Hannes Federrath\footnote{%
    Universität Hamburg, \email{hannes.federrath@uni-hamburg.de}
  }
}
\booktitle{GI SICHERHEIT 2022} 
\year{2022}

\maketitle

\begin{tikzpicture}[remember picture, overlay]
  \node[font=\sffamily\normalsize, yshift=-1cm, text centered, text width=\paperwidth, anchor=north west] at (current page.north west) {%
This is the authors' manuscript version.
  };
\end{tikzpicture}

\begin{abstract}
Case studies of application software data models indicate that timestamps are excessively used in connection with user activity.
This contradicts the principle of data minimisation which demands a limitation to data necessary for a given purpose.
Prior work has also identified common purposes of timestamps that can be realised by more privacy-preserving alternatives like counters and dates with purpose-oriented precision.
In this paper, we follow up by demonstrating the real-world applicability of those alternatives.
We design and implement three timestamp alternatives for the popular web development framework Django
and evaluate their practicality by replacing conventional timestamps in the project management application Taiga.
We find that our alternatives could be adopted without impairing the functionality of Taiga.
\end{abstract}


\section{Introduction}%
\label{sec:intro}

The design of software is today probably one of the biggest factors for everyday privacy.
Since using software becomes virtually inescapable,
it is increasingly application data modelling that decides how much of our personality and about our activities
is recorded.
Previous work~\cite{burkert_towards} indicates that data models make excessive use of timestamps,
the data type that adds the particularly sensitive temporal dimension to profiling.
Timestamps have been previously observed to fulfil various functions in programming that 
not even require temporal properties.
Instead, timestamps are frequently used for ordering or determining state
(\eg, maintain order in which attachments were added).
Function that can easily be achieved with less privacy-invasive alternatives.
But also in cases where their temporal functions like universal comparability are actually used (\eg, time a bug report was filed),
there appears to be room for a reduction of the typical second or even microsecond precision,
to precisions that correspond more with human perception and increase privacy.
Tackling the excessive use of timestamps in data models is a matter of raising awareness
but also of providing ready to use alternatives.
In this paper, we provide and evaluate a first such framework of timestamp alternatives.
In summary, we make the following contributions:
\begin{itemize}
  \item We design more privacy-preserving alternatives for common use cases of timestamp data types as identified by prior work.
  \item We validate the design applicability with a case study of the application \textit{Taiga}.
  \item We provide an implementation for the popular web application framework Django.
  \item We evaluate and demonstrate the practicality of those alternatives by replacing timestamps in data model of Taiga with our alternatives and observe the effects.
\end{itemize}

The remainder of the paper is structured as follows:
We firstly present related work and our adversary model,
then we describe the design and implementation of our alternatives,
after which we provide an evaluation.

\section{Related Work}%
\label{sec:relwork}

In a prior case study of the Mattermost application,
we systematically analysed the usage of personally identifiable timestamps in data models~\cite{burkert_towards}.
We found that timestamps of creation, last modification and deletion are included in a majority of models.
However, most user-related timestamps were found to have no programmatic use and only a small fraction is displayed on the user interface.
Based on the identified functions of timestamps, we proposed design alternatives that use precision reduction and context-aware counters.
Otherwise, the literature on timestamp-related privacy patterns and practical data minimisation is scarce.
In 2017, a literature survey of privacy patterns by \citet{DBLP:conf/euromicro/LenhardFH17} showed that proposals are rarely verified or even implemented.
Strategies to reduce the sensitivity of timestamps have been proposed by \citet{DBLP:conf/securecomm/ZhangBY06} for log sanitization.
They discuss \textit{time unit annihilation} as a strategy to gradually reduce precision over time.

\section{Adversary Model}%
\label{sec:attacker}

To contextualise privacy gain through our more data-minimal timestamp alternatives,
we provide the following adversary model.
It follows the established honest-but-curious~(HBC) notion commonly used to assess communication protocols.
\textcite{hbc} define an HBC adversary as a legitimate participant in a communication protocol, who will not deviate from the defined protocol but will attempt to learn all possible information from legitimately received messages.
Following the adaption of this model to the context of application software and data models~\cite{burkert_towards},
we consider an adversary to be an entity that is in full technical and organisational control of at least one component of a software system, e.g., the application server.
The adversary will not deviate from default software behaviour and its predefined configuration options,
but will attempt to learn all possible information about its users from the data available in the application.
This especially means that an adversary will not modify software to collect more or different data, or employ additional software to do so.
However, an adversary can access all data items that are collected and recorded by the software system irrespective of their exposure via GUIs or APIs.
We reason that this adversary model fits real world scenarios, because software operators in general lack the technical abilities to modify their software systems or are unwilling to do so, to not endanger the stability of their infrastructure
or to not document potentially illegal behaviour.
We come back to this adversary model when we employ server-side reduction later on.

\section{Design}%
\label{sec:design}

Based on alternative concepts for timestamps in the literature,
we designed three data types:
a generalised date with a static precision reduction~(\textit{\rd}),
a context-aware counter for chronological ordering~(\textit{\od}),
and a generalised date with temporally progressing precision reduction~(\textit{\vd}).
The designs are targeted as replacements for the conventional timestamp data type in the Django framework,
but are using only standard database features typically available in development frameworks.
The following describes the design for each alternative type.

\subsection{Type 1: Rough Date}%
\label{sec:rough}

Rough date is a variation of Django's standard \dtf\ that truncates the date to a given precision.
As such, it should offer all functionality that \dtf\ does and maintain the same interface,
to be usable as a drop-in replacement.
The desired precision is given as a mandatory argument at field initialisation, either in seconds
or in the style of the \lstinline|timedelta| class from Python's standard library package \lstinline|datetime|~\cite{py-datetime}
as multiples of the units minutes, hours, etc.
The following creates a \rd\ with one hour precision:
\lstinline|RoughDateField(hours=1)|.
We deliberately do not provide a default precision to force users to consider the necessary precision for their given use case.
The date part below a given precision is truncated.

\subsection{Type 2: Ordering Date}%
\label{sec:order}

\begin{figure}[htpb]
  \centering
  \includegraphics[width=.7\linewidth]{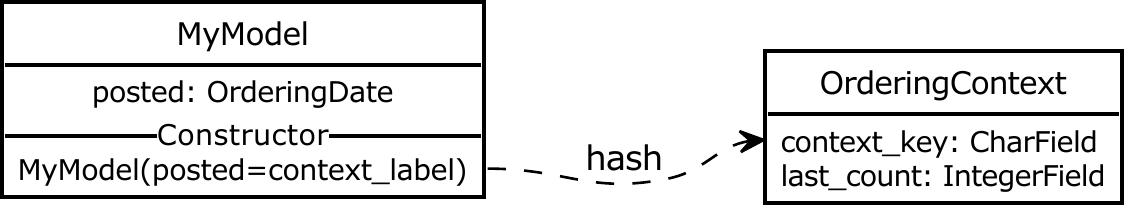}
  \caption{Class diagram showing a user-defined model with \odf. The related \con\ is identified by a context label given as initial field value. The integer value of \odf\ is then set to the context's next count.}%
  \label{fig:ordering-uml}
\end{figure}

The \od\ is an alternative to using timestamps for ordering items chronologically, if absolute date references and relative distances are not actually needed.
\odf\ offers ordering via context-specific auto-incremented counters.
Consequently, \od\ requires that objects are inserted in chronological order.
As shown in \cref{fig:ordering-uml}, a context-defining string key is given for each \odf\ at model initialisation.
The context label is cryptographically hashed to a \SI{256}{bit} key which then uniquely identifies its corresponding \con\ which persists the actual counter state information.
This way of maintaining the relation between \od{}s and their contexts in code and not in the database is space-efficient
and allows for dynamic contexts keys.
And since the context label is given at initialisation, ordering contexts can be defined very flexible.
For instance, a label can comprise a username and thereby create an isolation between counter contexts
of different users.
This can be used to increase user privacy by avoiding an otherwise global counter context that would make instances with \od\ temporally chronologically comparable across users.

\subsection{Type 3: Vanishing Date}%
\label{sec:vanish}

Vanishing date implements the privacy pattern of \textit{time unit annihilation}.
This alternative offers a progressing reduction of precision according to given increments until the end precision is reached.
For each step, a precision is provided like for \rdf\ in combination with a temporal offset, \ie, the distance from object creation that marks when the reduction step is due.
A background process regularly checks for due reductions and applies them.
\Cref{lst:vd-example} shows an example of a \vd\ with a three step reduction policy,
the first of which is immediately on creation,
whereas the second and third follow after a given time.
\Cref{tab:vanish-demo} lists the resulting stored date and next reduction event for each step.

\begin{lstlisting}[%
  language=python,
  caption={%
    Construction of a \vd\ with a three step reduction policy ranging from initially 1~hour to finally 1~month precision after 7~days.
    Helper \lstinline|make_policy| ensures correct reduction progression.
  },
  label=lst:vd-example,
]
created_at = VanishingDateField(policy=make_policy([
  Precision(hours=1),
  Precision(days=1, after_hours=3),
  Precision(months=1, after_days=7),
]))
\end{lstlisting}

\begin{table}[htb]
  \centering
  \begin{tabular}{@{}l*{3}l@{}}
  \toprule
   \textbf{Step} & \textbf{Current Time$^*$} & \textbf{Stored Date} & \textbf{Next Due Date} \\\midrule
   Creation/1st Red. & 2021-11-08 15:17 & 2021-11-08 15:00 & 2021-11-08 18:00\\
   2nd Reduction & 2021-11-08 18:01 & 2021-11-08 00:00 & 2021-11-15 00:00\\
   3rd Reduction & 2021-11-15 00:03 & 2021-11-01 00:00 & -\\
  \bottomrule
  \end{tabular}
  \caption{Exemplary progression of vanishing date reduction with a 3-step policy leading to a precision of one month after seven days.
  ($^*$Current times depend on the frequency and delay of periodic checks.)}%
  \label{tab:vanish-demo}
\end{table}

\begin{figure}[ht]
  \includegraphics[width=1\textwidth]{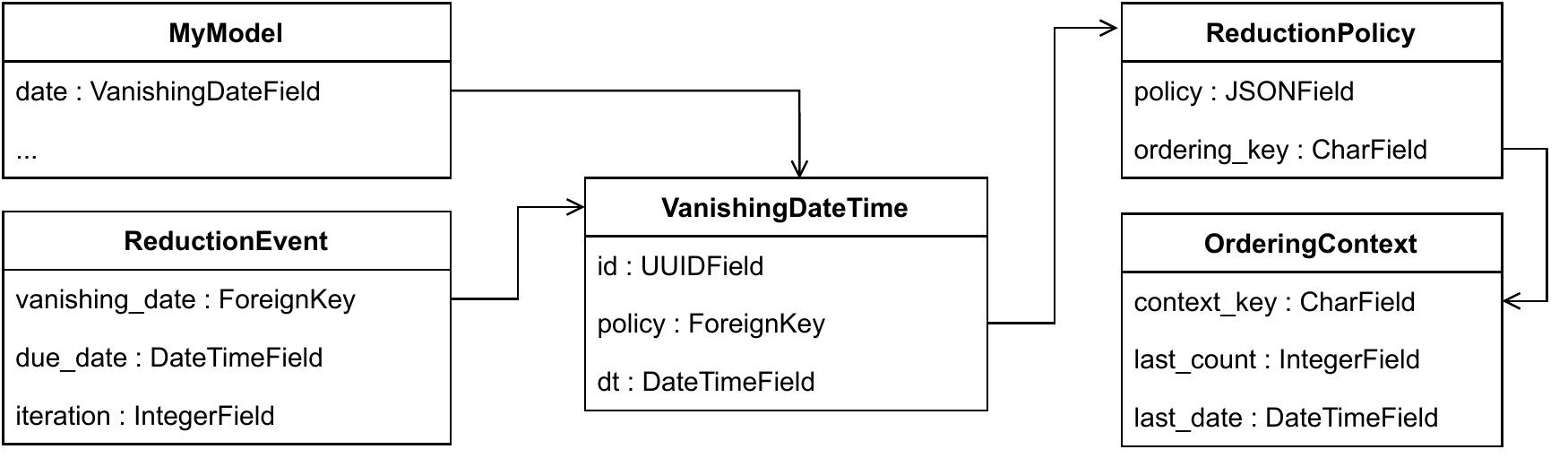}
  \caption{Class diagram showing a custom model that uses \vdf\ which references a \vdt\ object specifying the information about the reduction policy and events.}%
  \label{fig:vanish-uml}
\end{figure}

As shown in \cref{fig:vanish-uml}, the resulting design of \vd\ is more complex than for \rd\ and requires auxiliary models
to persist information about the reduction policy and the current progress within that policy.
Therefore, \vdf\ sets a reference to a \vdt\ class that holds the actual, gradually reduced date, a reference to a policy instance,
and to a \ve\ that represents the next due reduction step.
All instances of \ve\ form a queue that can be efficiently processed by the periodic due check.
Since Django lacks the ability to natively trigger periodic tasks,
we offer a management command for periodic reduction that can be triggered via, \eg, \textit{Cron}.

Note that to not leave any traces of the previously truncated time information,
due dates for subsequent reduction steps are calculated on the basis of the reduced step.
In \cref{tab:vanish-demo}, for instance, the second reduction is due at 18:00 instead of 18:17,
to not thwart the reduction to hour precision in the first step.
As a result, the time periods given between each policy step are upper boundaries.
In the previous example, the hour precision is available for 17~minutes less than the full three hours given in the policy.
Also note, the reduction level of a step should not be larger than the offset of the following step.

Following our adversary model in \cref{sec:attacker}, the application itself holds the only record of the timestamp reduced by \vd.
This especially means, that the software operator does not keep a mirror of the information before reduction or of earlier reduction levels.

\subsection{Design Validation}%
\label{sec:valid}
The purpose of design validation is to examine whether the design mainly based on previous case studies
also holds for the application we selected to evaluate our implementation.
As described below, we inspected its timestamp usage following the methodology of~\cite{burkert_towards}.

Taiga is a project management software that is built on the Django framework.
Taiga focuses on user interaction like the creation, processing and commenting of tasks to control and document the progress of projects.
Following the agile approach, interactions occur around planning elements like tasks, issues, epics, sprints and user stories.
We chose Taiga for our evaluation because it is a popular app built on Django and it is focused on structuring and recording user interactions,
which likely brings sufficiently complex requirements to test our implementation.
In the following, we describe our methodology, the identified timestamp use, and any necessary modifications to our design.

\subsubsection{Methodology}%
Following~\cite{burkert_towards},
we examine the source code of Taiga's back-end component~\cite{taiga-back-6.0.7} for occurrences of Django's date-related model fields
\lstinline|DateTimeField|, \lstinline|DateField|, and \lstinline|TimeField|.
We assess semantic and purposes of each timestamp by examining all of their uses
by the back-end, their presentation in the front-end~\cite{taiga-front-6.0.7} and their exposure via the API\@.
We then use the identified purposes to select from our proposed types an alternative that provides the required functionality.
If none should be available, our design would need refinement.
We find that the back-end REST API typically returns every attribute (model field) related to the requested object,
regardless of whether the front-end uses them or not.
Therefore, it is not sufficient to access timestamp usage and purpose only based on API exposure,
but actual use based on inspections of the rendered front-end are necessary.
To do so, we manually examined Taiga's front-end user interface cataloguing presented timestamp information.
Usage in the back-end was assessed by manually inspecting all occurrences of date-related field names throughout the back-end code.
These analyses were conducted on version 6.0.7 of both back-end and front-end, as released on March 8th, 2021.

\begin{table}\centering
  \begin{tabular}{@{}lrrr@{}}\toprule
    \textbf{Model Field} & \textbf{Occurrences} & \textbf{Used in Models} & \textbf{Exposed via API} \\\midrule
    DateTimeField & 170 & 48 & 41 \\
    DateField & 15 & 3 & 3 \\
    TimeField & 5 & 0 & 0\\
    \bottomrule
  \end{tabular}
  \caption{Usage of date-related Django model fields in Taiga's back-end.}%
  \label{tab:taiga-fieldstats}
\end{table}

\subsubsection{Identified Timestamps}%
\Cref{tab:taiga-fieldstats} shows the numbers of identified uses per model field.
In total, we located 190~occurrences in Taiga's back-end code of which 51 are part of data model definitions.
The remaining matches occurred in database migrations and serialization code.
We did not further inspect the latter occurrences as they do not contribute any usage and purpose information.
Almost all definitions use the \lstinline|DateTimeField|.
Only 3 (\SI{6}{\percent}) use the \lstinline|DateField|, whereas \lstinline|TimeField| is not used in current model definitions at all.

To assess the API exposure, we inspected the source code and consulted the official API documentation for information about which fields are included in a query response.
We found that all but 7~timestamps (\SI{86}{\percent}) are exposed through the API\@.
Of those 7 timestamps, 5 are also not programmatically used on the back-end.
The visual inspection of the front-end UI also revealed that at least 22 (\SI{50}{\percent}) of the timestamps fetched from the back-end API are not used there.
Regarding those timestamps without a detectable usage or purpose,
we can not determine a purpose-appropriate alternative.
For the sake of data minimisation, they should be removed entirely.
Also, not all timestamp fields are necessarily personal data.
This is true for the three \lstinline|DateField| uses, which are used to model due dates of planning elements (\eg, sprints) which are not directly linked to actions of users.
Hence, we omit those from classification as well.
For the remaining timestamps, we classified their type purpose according to their usage context in back-end and/or UI\@.

\subsubsection{Timestamp Semantic and Purpose Classification}%

We classified the remaining timestamps based on the semantic given by their variable name and source code context.
All models in Taiga~(27) have a creation timestamp to automatically capture when a model instance was created.
17~models additionally record the time of the latest update,
three the time a planning element was completed, and one when a notification was read.
Regarding purpose classification, we followed~\cite{burkert_towards} and used a bottom-up classification
that inspects and labels each timestamp's programmatic use in the source code
with respect the function they serve in the respective context,
resulting into similar purposes:
presentation for user information, sorting, and comparison.
\Cref{tab:taige-ts} in the appendix lists the identified purposes for each timestamp.

\subsubsection{Design Revision}%
Based on the identified purposes and functional properties of our proposed alternatives,
we select possible replacements for each timestamp.
The selections are shown in \cref{tab:taige-ts}~(appendix).
If a timestamp is only used for sorting like the creation date of \texttt{Attachment}, the \od\ is the apparent alternative.
Timestamps with a presentation or comparison purpose can equally be replaced by \rd\ and \vd.
The latter should be chosen if an initial higher demand for precision exists.

We find that our proposed alternatives cover all found purposes.
However, we noticed that the purpose of maintaining temporal order sometimes coincides with providing a temporal context (presentation or comparison).
To replace such a timestamp,
two fields are required in the initial design~(e.\,g., \odf\ and \vdf).
Since this would both complicate usage and increase memory footprint,
we decided to introduce two additional fields that combine the properties of \od\ with \rd\ and \vd\ respectively,
which otherwise do not maintain order for dates reduced to the same value.
To do so without increasing memory footprint,
we use the sub-second value range available in most timestamp representations
to hold the ordering counter.
For a microsecond timestamp this leaves space for a \(10^6\) counter.
The counter is incremented for all timestamps with identical values in their end-precision~(\cref{tab:ts-samples}).
Note that this approach only works for timestamps that are added in chronological order, e.\,g., that are automatically set to the current time, otherwise the insertion order would not reflect their temporal order.

\begin{table}\centering
  \begin{tabular}{@{}lccc@{}}\toprule
     \textbf{Original} & \textbf{Vanishing} & \textbf{Vanishing+Order} & \textbf{Vanishing+Order} \\
     & 1. Iteration [\SI{5}{\sec}] & 1. Iteration [\SI{5}{\sec}] & 2. Iteration [\SI{30}{\sec}]\\\midrule
     12:20:11:673320 &  12:20:10:000000 & 12:20:10:000000 & 12:20:00:000000 \\
     12:20:14:313406 &  12:20:10:000000 & 12:20:10:000001 & 12:20:00:000001 \\
     12:20:17:248323 &  12:20:15:000000 & \textbf{12:20:15:000002} & \textbf{12:20:00:000002}  \\
     12:20:33:040852 &  12:20:30:000000 & 12:20:30:000000 & 12:20:30:000000 \\
     12:20:35:917632 &  12:20:35:000000 & 12:20:35:000001 & 12:20:30:000001 \\
    \bottomrule
  \end{tabular}
  \caption{Sample sequence of \vd\ with and without added support to preserve ordering.
  The highlighted timestamps demonstrate that counter reset is determined by the end-precision of \SI{30}{\sec} which defines the counter scope.
  }%
  \label{tab:ts-samples}
\end{table}

\section{Implementation}%
\label{sec:impl}

We implemented our revised concept as a Django app that can be included into other Django projects to provide our date alternatives.
It is available open source on GitHub~\cite{privacy_dates_src}.
In the following, we describe trade-offs and limitations of this implementation.
Ordering date can simply build on available counter fields and is hence omitted from description.

\subsection{Rough Date}%

To offer \rdf\ as a drop-in alternative for \dtf,
it also has to support the options \lstinline|auto_now| and \lstinline|auto_now_add|,
which automatically set the field to the current time at the moment when the object is saved (not initialised).
To support these options, the reduction of precision has to be integrated in the saving process,
since at any earlier point the value is not yet defined.
To do so, we use pre-save hooks that apply the reduction.
As a consequence, date values assigned to \rdf\ remain in full precision until saved.

\subsection{Vanishing Date}%

As \vd\ is the most complex type, we face three main implementation challenges.

\paragraph{Avoiding chronology leak with UUIDs}%
By default, Django would use an auto-incrementing integer primary key for \vdt\ if no other primary key was specified.
Auto-incremented integers would however leak information about the temporal creation order of all instances of every model that uses \vdf.
For instance, an attacker could learn that user A logged in after user B posted their last comment but before user B closed the issue.
To prevent such chronology leaks, we use randomized UUIDs as primary key.
It should be noted that databases might still leak the temporal order by exposing the insertion order in certain queries.
Future work should investigate options to, \eg, prevent users from executing such ordering queries.

\paragraph{Policy Reuseability}%
As previously shown in \cref{fig:vanish-uml}, we decided to make \vp\ a separate model to allow its reuse among \vd{}s with the same policy.
We provide the helper function \lstinline|make_policy| that transparently ensures policy reuse.

\paragraph{Model Identification with \mix}%
An identification of all models that make use of \vdf\ is required for two reasons:
Firstly, to ensure a two-way cascading delete of \vdt,
and secondly to create an initial \ve.
To locate the relevant models, we decided to employ the common mix-in pattern,
which uses inheritance to add functionality to a given class or model.
Users of \vdf\ have to add the \mix\ as a base class for their model.
Thereby, we can automatically find all sub-classing models and register post-delete listeners to ensure a two-way deletion,
as well as a post-safe listener to update reduction events.

\section{Evaluation}%
\label{sec:eval}

In this section, we evaluate the practicality of our framework and its storage cost.

\subsection{Practicality of Taiga Integration}%
\label{sec:eval:taiga}

To evaluate the practicality of our alternatives,
we modified Taiga version~6.0.7 to use the timestamp replacements identified in \cref{sec:valid} and detailed in the appendix.
To test the correctness and impact of our modifications, we ran the modified Taiga and inspected the error output as well as the web front-end for potential negative effects.
To check functional integrity, we created and modified various planning elements and compared the visible front-end behaviour before and after integrating our alternatives.

We found that all timestamps could be replaced as suggested, with few adjustments to the code base.
As expected, \rd\ required the least effort of only replacing the field type.
More effort was needed for the other alternatives:
We used \vd\ and \od\ to each replace timestamps in three models.
Since both replacements do not behave like a standard \dtf,
all code that accessed or modified the field value had to be adjusted to either use the reference \vdt,
or an appropriate context label instead.
After these modifications, Taiga operated normally and we were able to create and modify elements as usual
without functional impairments visible through UI or error log.

When it comes to presenting the replacements in the UI,
\rd\ and \vd\ can mostly be treated like standard dates.
However, to avoid confusion or wrong expectation of precision,
the formatting of both should be adopted to reflect their precision.
In contrast, \od\ can no longer be presented as a date in a meaningful way.
But if the timestamp previously only fulfilled ordering purposes this should not be an issue.
Otherwise, \vd\ might be a more fitting replacement.

\subsection{Storage Cost}
The following assesses storage cost compared to ordinary date and time~(\SI{8}{byte}) using MariaDB as example~\cite{MariaDB2019}.
Rough date has the same memory footprint.
Ordering date uses a \SI{4}{byte} counter reducing cost by half.
The cost of \vd\ is dominated by three UUIDs, which 
require \SI{38}{byte}.
Added to two \SI{8}{byte} date, one foreign key and one event counter~(each \SI{4}{byte}),
a total of \SI{138}{byte} is required for \vd, which is 17.25 times the cost of an ordinary date and time.
Additionally, usage-dependent storage cost is added for \vd\ and \od\ by their auxiliary models.
The number \con\ used depends on the number of distinct context labels set by developers,
and scales with the number of users if individual contexts were used to avoid unnecessary comparability.
In MariaDB, each \con\ requires 44 bytes.
Moreover, a variable amount of storage is required for \vp, which depends on the number of defined reduction steps.
To give an overall example, applying the replacements in \cref{tab:taige-ts} increases Taiga's average storage cost per timestamp by about 5 times (not weighted by instance frequency).

\section{Conclusion}%
\label{sec:concl}

Excessive, unthought use of timestamps in software data models is a violation of the data minimisation principle and potentially harmful to user privacy.
Our case study of the Taiga application not only supports the findings of prior work regarding excessive use,
but also in terms of minimisation potential through the use of more-privacy preserving timestamp alternatives.
We have presented a framework of alternatives for common timestamp functions and purposes.
We demonstrated its practicality by implementing it as a Django app which we then used to replace timestamps in Taiga.
Although demonstrated for Django, these alternatives only use standard concepts and can be implemented for other development frameworks.
Our evaluation with Taiga revealed that code changes were necessary but limited to adopting changed initialisation and access methods.
Additionally, the presentation of timestamp may need adjustment to convey decreased precision levels.
Depending on the selection of alternatives, especially the frequency of \vd, storage cost might increase noticeably.
Where this is nor acceptable, \rd\ and \od\ can be used with little to no additional storage cost, but without gradual reduction.
Our integration test suggests that more privacy-preserving alternatives can be adopted with reasonably low effort.
A user study to evaluate their usability with developers is left to future work.

\paragraph{Acknowledgements}
The work is supported by the German Federal Ministry of Education and Research~(BMBF) as part of the project Employee Privacy in Development and Operations (EMPRI-DEVOPS) under grant~16KIS0922K.

\printbibliography%

\newpage
\appendix

\section{Taiga Timestamp Purposes and Replacements}%
\begin{table}[htp]%
  \footnotesize
  \newcommand\model[1]{\rowcolor{Lightgray}\multicolumn{5}{l}{\textbf{#1}}}
  \newcommand\y{\faCheck}
  \centering
  \begin{tabular}{@{}lcccc@{}}\toprule
    \textbf{Model} / Timestamp  & Presentation & Sorting & Comparison & Replacement\\  \midrule
    \model{Attachment}\\
    created\_date  &  & \y &  & OD \\
    \model{Epic}\\
    created\_date  &  \y & &  & RD \\
    \model{HistoryChangeNotification}\\
    updated\_datetime  &   &  & \y & RD \\
    \model{HistoryEntry}\\
    created\_at  &  \y & \y &  & VD+O \\
    delete\_comment\_date  &  \y &  &  & VD \\
    edit\_comment\_date  &  \y &  &  & VD \\
    \model{Issue}\\
    created\_date  &  \y &  &  & RD \\
    modified\_date  &  \y & \y & & RD \\
    \model{Like}\\
    created\_date  &   &  & \y & RD \\
    \model{Task}\\
    created\_date  &  \y &  &  & RD \\
    \model{TimeLine}\\
    created  &  \y & \y & \y & VD+O \\
    \model{User}\\
    date\_joined  &  \y &  &  & RD \\
    \model{UserStory}\\
    created\_date  &  \y &  &  & RD \\
    \model{Watched}\\
    created\_date  &   & \y &  & OD \\
    \model{WebNotification}\\
    created  &  \y & \y &  & VD+O \\
    read  &   & \y &  & OD \\
    \model{WikiPage}\\
    created\_date  &  \y &  &  & RD \\
    modified\_date  &  \y &  &  & RD \\
    \bottomrule
  \end{tabular}
  \begin{tabular}{lclclc}
    Rough Date \textbf{(RD)}& ~  &
    Ordering Date \textbf{(OD)}& ~  &
    Vanishing Date \textbf{(VD)}  \\
    \multicolumn{5}{c}{Vanishing Date with Ordering  \textbf{(VD+O)}}  \\
  \end{tabular}
  \caption{Identified purposes and suggested replacements for used timestamps in Taiga.}%
  \label{tab:taige-ts}
\end{table}

\end{document}